\documentclass{webofc}
\usepackage[varg]{txfonts}   
\usepackage{booktabs}
\usepackage{graphicx}
\usepackage{caption}
\usepackage{physics}
\usepackage{xcolor}
\usepackage{hyperref}
\usepackage{xr}
\begin{document}
\title{Benchmarking machine learning models for quantum state classification}
%
%

\author{\firstname{Edoardo} \lastname{Pedicillo}\inst{1,2}\fnsep\thanks{\email{edoardo.pedicillo@unimi.it}}
        \and
        \firstname{Andrea} \lastname{Pasquale}\inst{1,2}\fnsep\thanks{\email{andrea.pasquale@unimi.it}} \and
        \firstname{Stefano} \lastname{Carrazza}\inst{1,2,3}\fnsep\thanks{\email{stefano.carrazza@unimi.it}}
}

\institute{TIF Lab, Dipartimento di Fisica, Universit\`a degli Studi di
  Milano and INFN Sezione di Milano, Milan, Italy.
\and
           Quantum Research Center, Technology Innovation Institute, Abu Dhabi, UAE.
\and CERN, Theoretical Physics Department, CH-1211
  Geneva 23, Switzerland.
       }

\abstract{%
    Quantum computing is a growing field where the information is processed by two-levels
    quantum states known as \emph{qubits}. Current physical realizations of qubits require
    a careful calibration, composed by different experiments, due to noise and decoherence
    phenomena.
    Among the different characterization experiments, a crucial step is to
    develop a model to classify the measured state by discriminating the ground
    state from the excited state.
    In this proceedings we benchmark multiple classification techniques applied
    to real quantum devices.
}
\maketitle
\section{Introduction}

During the last decade, we have witnessed the spread of quantum devices granting users the
possibility to start deploying quantum algorithms on quantum hardware.
The typical way of accessing quantum processing units (QPUs) nowadays is through cloud
providers. Although this approach is user-friendly, it tends to hide complications related
to the deployment and maintenance of self-hosted QPUs, which are now accessible to research
institutions.

In order to fill this gap, middleware open-source frameworks are fundamental to
accelerate the growth and the development of research in quantum computing.
An example of such frameworks is \texttt{Qibo}~\cite{Efthymiou_2021,stavros_efthymiou_2023_7992830},
a full-stack quantum computing library providing an Application Programming Interface (API)
to build quantum computing algorithms based on circuit and adiabatic paradigms.
The latest addition to the Qibo framework are two modules dedicated to hardware
control and characterization: \texttt{Qibolab}~\cite{efthymiou2023qibolab,
stavros_efthymiou_2023_7973899} and
\texttt{Qibocal}~\cite{pasquale2023opensource, andrea_pasquale_2023_7957542}.

Hardware control is not sufficient to deploy algorithms on quantum hardware.
Current quantum technologies, especially superconducting devices, require a detailed
characterization of the experimental setup, which is achieved by performing several
experiments~\cite{PRXQuantum.2.040202} aim at fine-tuning specific parameters.
The need for calibration lies in the fact that superconducting qubits suffer from
both noise and decoherence \cite{Schlosshauer_2019} phenomena which can lead to a
shift in the characterization parameters. To address this issue there have been some
works in order to automate the calibration process using direct acyclic graphs
\cite{kelly2018physical} and optimization techniques.
In this context, several calibration libraries have appeared over the last years
\cite{kelly2018physical,Kanazawa2023,forestbenchmarking}, including \texttt{Qibocal}.

Among the different experiments required to achieve high-fidelity operations,
we are going to focus on the protocol which trains a Machine Learning (ML) model to
classify the measured state of a qubit.
In the case of an ideal system, it should be quite straightforward to discriminate the state
$\ket{0}$ from the state $\ket{1}$. However, this task can become non-trivial due to decoherence
phenomena which tend to produce errors. This is why it is necessary to train a ML
model to perform correctly the classification.

Such classification can be performed directly on the instrument which is collecting
the measurements or can be executed afterwards as a post-processing operation. Although the first
approach is generally faster, some devices have limited capabilities when it comes to
storing and training ML models. This is why they may fall back on simplified
models which can have worse performances.
On the other hand, performing the classification outside the instrument can
lead to better performances, but it may introduce some overhead.

In this proceedings we investigate this problem by benchmarking several ML classifiers
trained on data obtained by real quantum devices.

\section{Methods}

\subsection{Measuring a qubit}
In this section we are going to briefly present how qubits are measured in the case of superconducting
devices. The most known technology to realize superconducting qubits are transmons~\cite{PhysRevA.76.042319},
weakly anharmonic oscillators made using Josephson junctions~\cite{Josephson:1962zz}.

Transmons are measured by dispersively coupling them with resonators, also known as \emph{dispersive readout}.
It can be shown that in the non-resonant limit $|\Delta| \gg g$ the qubit-resonator system can be described
by the following effective Hamiltonian \cite{PhysRevA.69.062320,article}:

\begin{equation}
    \label{eq:spectroscopy}
    H \approx \hbar \underbrace{\bigg( \omega_r + \frac{g^2}{\Delta} \sigma_z \bigg)}_{w} a^{\dagger} a +
    \frac{1}{2} \hbar \bigg(\omega_a + \frac{g^2}{\Delta}\bigg) \sigma_z  \\,
\end{equation}
where $\Delta$ is the detuning between
the frequency of the resonator $\omega_r$ and the qubit frequency $\omega_a$ and $g$ is the resonator-qubit coupling.
The transition frequency of the resonator $w$ is now affected by the state of
the qubit:
\begin{equation}
    w =\begin{cases}
        w_r + \frac{g^2}{\Delta}, & \text{if $\ket{0}$}\\
        w_r - \frac{g^2}{\Delta}, & \text{if $\ket{1}$}
     \end{cases}\quad .
\end{equation}

We can detect this shift, and therefore identify the qubit state, by sending a readout probe signal
generated using Arbitrary Waveform Generators (AWGs). We can extract the reflected amplitude and
the phase of the microwave signal against the probe frequency $\omega$.
To properly observe the frequency shift we need to fine-tune the beam power to reach the
dispersive-regime.

\begin{figure}
    \includegraphics[width=.33\textwidth]{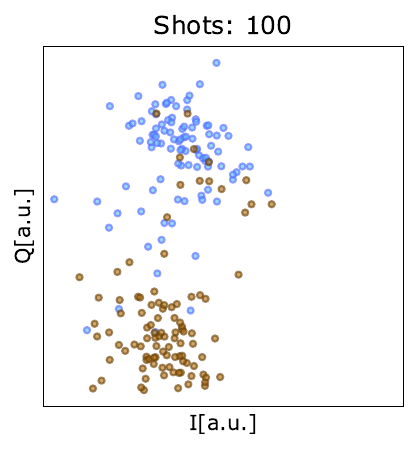}\hfill
    \includegraphics[width=.33\textwidth]{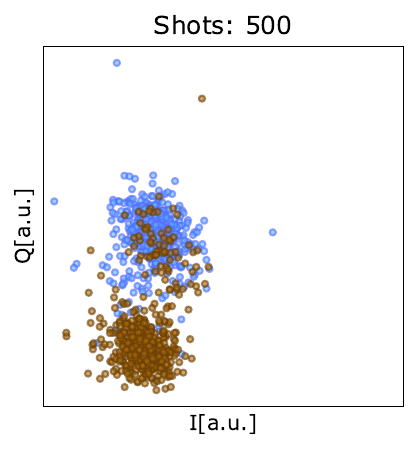}\hfill
    \includegraphics[width=.33\textwidth]{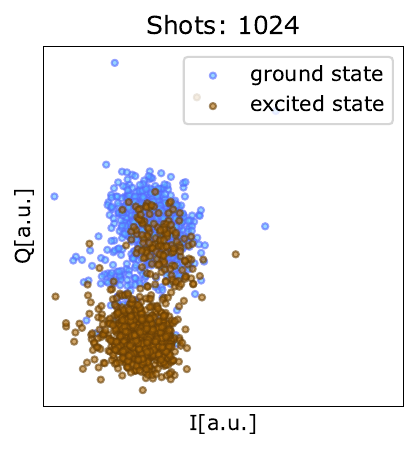}
    \caption{$IQ$ components for different shots acquired by a real quantum device.}\label{fig:clouds}
\end{figure}

Another way to visualize the population of the ground and excited state is by demodulating
the microwave probe into its in-phase $I$ and quadrature component $Q$. By demodulating we are
projecting the signal from the waveform space onto a 2D plane with coordinates $I$ and $Q$.
The ground and the excited states are identified by the two clouds shown in
Fig.~\ref{fig:clouds}. This distribution is caused by decoherence and noise mostly coming
from amplifiers \cite{Arute_2019}.

We can expect that a generic measurement will generate some points in the $IQ$ plane and we are
interested in identifying if the point observed corresponds to $\ket{0}$ or $\ket{1}$.
This problem is well known in ML and it is often referred to as \emph{classification} or
\emph{discrimination}~\cite{Gareth}.
There are several models which can tackle this problem as we will show shortly.

This classification experiment is usually performed after the calibration of the $\pi$-pulse,
which aims at fine-tuning the parameters of a drive pulse to produce the excited state $\ket{1}$.
The raw data are generated using two different sequences of pulses. We produce a series of 0s
by simply measuring the state of the qubit, while the 1s are producing by measuring the qubit
after sending a $\pi$-pulse.
For each shot we store both the $I$ and the $Q$ component.
As we can see in Fig.~\ref{fig:clouds}, the distribution becomes more complicated as the number of shots increases. We can
also observe that some excited states decay into the ground state due to short lifetime of the qubit.
Finally, the data are fed into a classification model which will be trained accordingly.

\subsection{Classification models}

It follows a brief description of the models that we have trained, including
some of their strengths and weaknesses.

\subsection*{Linear SVM}

A support vector machine (SVM) is a collection of supervised learning algorithms, that can be used both for classification
and regression.
Its versatility comes from the possibility of using different kernels, the most used is the linear kernel.
As a classifier, the linear SVM is trained in order to find the best hyperplane that separates the two classes.
The advantages of using this model are its robustness against overfitting due
to its regularization parameter and its memory efficacy, since it uses a subset
of training points, called \emph{support vectors}, in the decision function.
It is suitable when the number of dimensions exceeds the sample's one.
On the other hand, it is sensitive to outliers that might affect the separation hyperplane,
it is computationally expensive for large datasets and does not directly provide probability estimates.

\subsection*{AdaBoost}

The main idea behind the AdaBoost algorithm is to generate and train a sequence
of models, usually decision trees, that are slightly better than random guessing,
usually called \emph{weak learners}.
Given a new input, the model classifies it by collecting the output of each weak learner
and combining them with suitable weights.
Using AdaBoost, it is less likely that overfitting will occur since it uses an ensemble of
weak learners that underfits the training set. The convergence is guaranteed,
since the training error decreases exponentially with the number of epochs.
However, it is sensitive to noisy data, which can negatively impact performance,
and computationally more expensive than other algorithms.
The training of the different learners cannot be parallelized, unlike random forests.

\subsection*{Gaussian Process}

A Gaussian process classifier works by generating a function, \emph{nuisance function}, $f(x)$ from
a Gaussian process, and then squashing it through a function $\sigma$ called link function
(usually the logistic logit function). Therefore, the class probability is $\sigma(f(x))$.
The values of $f$ are not particularly relevant for the model,
since the probability of belonging to a class is given by the average of class probabilities
over the nuisance function space. Although the logistic link function integral cannot be calculated analytically,
it can be easily approximated in the binary case.
On the other side, it could be computationally demanding and may not scale efficiently 
to large datasets due to the lack of sparsity.

\subsection*{Naive Bayes}

Naive Bayes methods are a set of supervised learning algorithms based on applying
Bayes’ theorem with the \emph{naive} assumption of conditional independence between
every pair of features given the value of the class variable.
It usually requires a small amount of training data, and the features' independence 
Ansatz could solve problems related to the curse of dimensionality.
It could handle missing values and irrelevant features effectively.
In the worst-case scenario, the independence assumption may be incorrect,
or outliers and imbalanced datasets may reduce the model's effectiveness.

\subsection*{Neural Network}

A neural network is a collection of connected nodes, each of which performs a
specific operation on the input data.
The connection topology and operation of the nodes differ between neural network models.
We use a Feedforward Neural Network (FNN) in this proceedings, which is distinguished
by one-way information transmission and the division of nodes into layers that
are fully connected with the previous and next ones.
When a non-linear approach is required, the FNN can be used for both regression and classification.
Neural Networks have numerous advantages, including the ability to learn complex
patterns and relationships in the training dataset, extreme flexibility, and good performance
on large datasets with high dimensionality.
The careful tuning of the hyperparameters to avoid overfitting and the training
could require high computational costs.
Because the loss function is non-convex, there can be several local minima, which
suggests that the model is sensitive to weight initialization.

\subsection*{Random Forest}

The Random Forest is a collection of trees that are generated by selecting a random
sample of the training set with replacement. The best split of each tree could be
evaluated either from all features or a random subset.
This bootstrapping technique reduces model variance and the risk of overfitting,
but it may slightly increase bias.
The model predicts the class by averaging the output of each tree, which cancels
out some errors.
Although it is possible to parallelize tree construction and predictions,
it may be computationally expensive for large datasets during training.

\subsection*{Radial Basis Function Support Vector Machine}

Support Vector Machine with radial basis function~\footnote{
Given  $\bold{x}$ and $\bold{x'}$ as two feature vectors in the input space
    the radial basis function is defined as
\begin{equation*}
    K(\bold{x}, \bold{x'}) = exp(- \gamma \norm{\bold{x}-\bold{x'}}^2).
\end{equation*}
} as kernel.
It is effective for non-linearly separable data through the use of kernel functions,
and it has only two hyperparameters that require careful selection and tuning.

\subsection*{Fidelity fit}

This method \cite{Reed_2010} evaluates the axis which passes through the centroids of the two clusters
in the training set, and determines the optimal threshold that maximizes the difference
between the cumulative distributions of the projections along the axis.
Predicting the state of a new point involves determining its relative position
to the threshold based on its projection along the axis.
Due to its small complexity, it has high interpretability, but it does not evaluate the probabilities.

\subsection{Training}

The initial step involved splitting the dataset into a test set, containing 25\% of
the data, and a training set. Following this, we optimize the hyperparameters for each classifier whenever possible.

The neural network was initialized by the \texttt{Keras}~\cite{chollet2015keras} library, its hyperparameters optimization
was carried out by a variant of the \texttt{HyperBand}~\cite{hyperband} algorithm provided by \texttt{Keras Tuner},
the fidelity fit model was implemented from scratch in \texttt{Qibocal}, and the other models
are those available in the \texttt{scikit-learn}~\cite{scikit-learn} library, their best hyperparameters
are found using a grid search algorithm.
The hyperparameter optimization details are tabulated in Tab.~\ref{tab:hyperparameters}.

\begin{table}
    \begin{tabular}{llc}
        \toprule
        \textbf{Model}          & \textbf{Hyperparameters}                       & \textbf{Configuration space}                                               \\
        \midrule
        Ada Boost      & estimator number                & {[}10, 200{]}                                      \\
                    & learning rate                   & {[}0.1, 1{]}                                     \\
                    & algorithm                       & \textit{SAMME}, \textit{SAMME.R}                                 \\
        \midrule
        Neural Network & size first layer                & 16, 1056                                   \\
                    & size second layer               & 16, 1056                                   \\
                    & activation function             & \textit{relu}, \textit{sigmoid}, \textit{tanh}, \textit{RBF}                       \\
                    & learning rate                   & $10^{-4}$, $10^{-2}$ \\
                    & optimizers                      & \textit{Adam}, \textit{Adagrad}, \textit{SGD}, \textit{RMSprop}                    \\
        \midrule
        Random Forest  & estimators number               & {[}10, 200{]}                                      \\
                       & criterion                       & \textit{gini}, \textit{entropy}, \textit{log\_loss}                       \\
                       & number of feature for splitting &\textit{sqrt}, \textit{log2}, \textit{all}                                \\
        \midrule
        RBF SVM        & regularization parameter      & {[}0.01, 2{]}                                      \\
                    & degree                          & 2, 3, 4 \\
        \bottomrule
    \end{tabular}
    \caption{Table listing all the hyperparameters that were optimized and their corresponding values.}
    \label{tab:hyperparameters}
\end{table}

\section{Results}
\begin{table}
    \resizebox{\textwidth}{!}{
    \centering
    \begin{tabular}{lccccc}
        \toprule
        \textbf{Name} &  \textbf{Accuracy} &  \textbf{Test Time ($\mu$s)} &  \textbf{Train Time (s)} & \textbf{True Positive Rate} & \textbf{False Positive Rate} \\
        \midrule
        Ada Boost &  0.904 &  16.22 &       0.401 & 0.94 & 0.13 \\
        Linear SVM &  0.905 &  13.13 &       0.40 & 0.94 & 0.12 \\

        Gaussian Process &  0.904 &  78.97 &     319.728 & 0.93 & 0.12\\
        Naive Bayes &  0.901 &  0.57 &       0.005 & 0.92 & 0.12\\
        Fidelity Fit   &  0.904 &  0.07 &       0.129 & 0.95 & 0.14\\
        Random Forest &  0.890 &  24.47 &       0.720 & 0.91 & 0.12\\
        RBF SVM &  0.902 &  54.29 &       0.163 & 0.94 & 0.13\\
        Neural Network &  0.903 &  132.93 &      59.793 & 0.95 & 0.15\\
        \bottomrule
        \end{tabular}
    }
    \caption{Table collecting the different parameters that we took in
    consideration to benchmark the different models.}
    \label{tab:benchmarks}
    \end{table}

\begin{figure}
    \includegraphics[width=.33\textwidth]{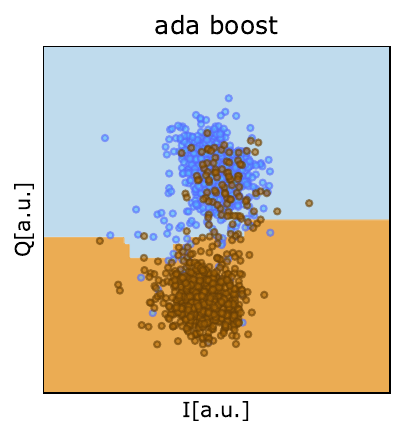}\hfill
    \includegraphics[width=.33\textwidth]{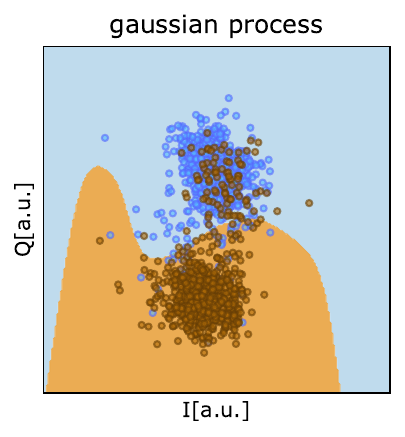}\hfill
    \includegraphics[width=.33\textwidth]{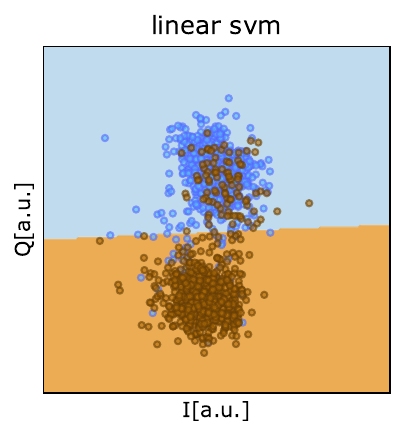}
    \\[\smallskipamount]
    \includegraphics[width=.33\textwidth]{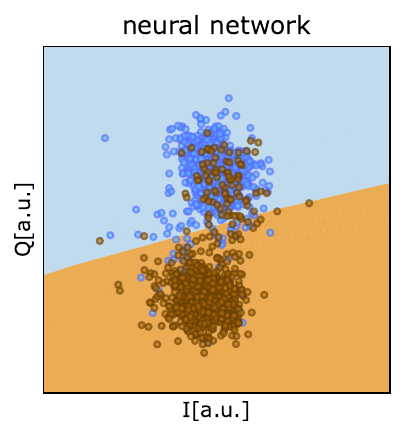}\hfill
    \includegraphics[width=.33\textwidth]{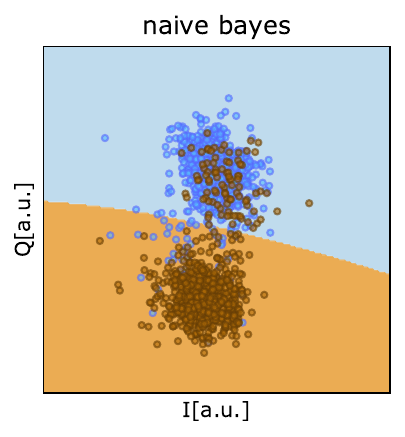}\hfill
    \includegraphics[width=.33\textwidth]{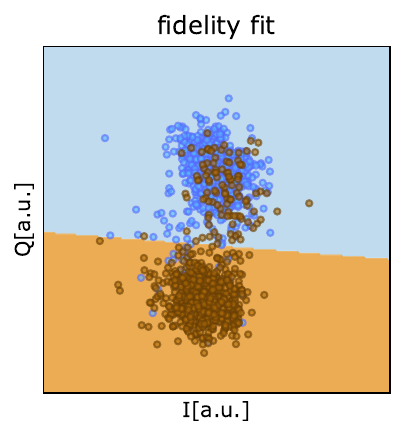}
    \\[\smallskipamount]
    \includegraphics[width=.33\textwidth]{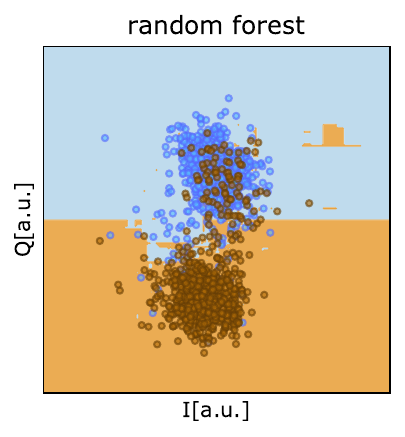}
    \includegraphics[width=.33\textwidth]{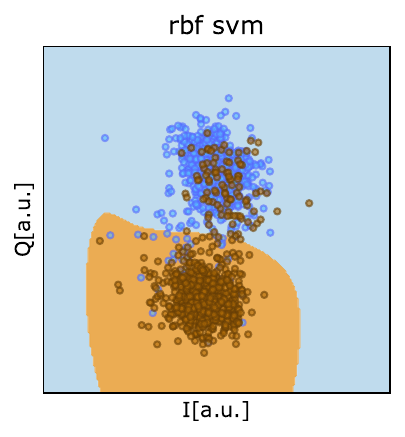}\hfill
    \caption{Classifiers' predictions (background colors) with test set (scattered points). }\label{fig:predictions}

\end{figure}

\begin{figure}
    \centering
    \includegraphics[width= .7\textwidth]{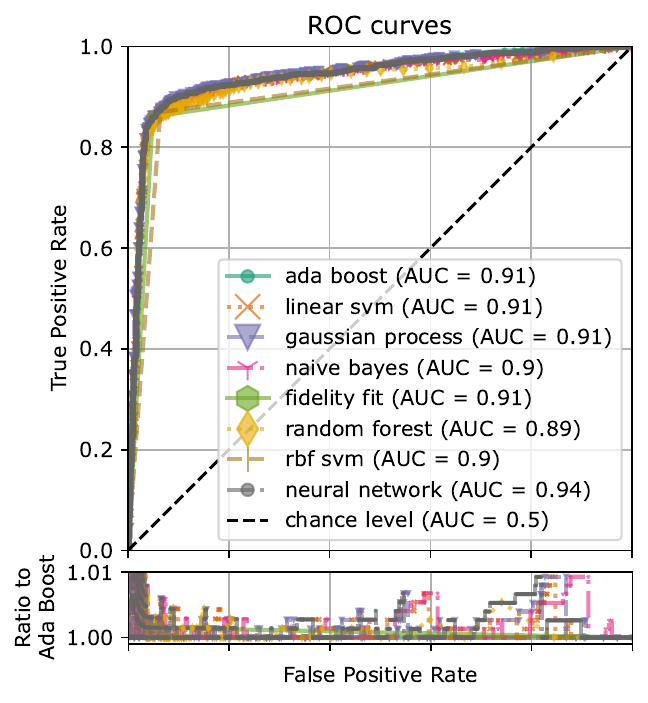}
    \caption{ROC curves computed for each classifiers
    (\emph{top}) and ratio w.r.t. Ada Boost classifier (\emph{bottom}).
    }\label{fig:ROC}

\end{figure}

A visualization of the classifiers' prediction capability is shown in Fig.~\ref{fig:predictions}. The scattered points
represent the testing sample while we show in the background the prediction for each model.
All models seem to have comparable performances as shown in Tab.~\ref{tab:benchmarks}
with accuracies distributed in a narrow interval around 0.90.
Looking at the ROC curves in Fig.~\ref{fig:ROC} we can conclude that the performances
between all the models are similar with a slight advantage of the Neural Network
that presents the best AUC.
Given that all the models perform similarly in terms of all the quantities that describe the
quality of a classifier (accuracy, true positive rate, false positive rate, AUC),
selecting the optimal one is not trivial.

All the classifiers have testing times that are reasonably fast, ranging from
0.07 to 132.93 micro-seconds. These significant variations in testing time suggest
that choosing the wrong classifier may generate an overhead which scale with the number of shots.
This could affect several quantum algorithm such as variational quantum circuits where
the training requires generating thousands of shots in each epoch.
Also, the length of training period differs substantially. In comparison to the other
classifiers, Gaussian Process and Neural Network have much longer training periods.
On the other hand,  Naive Bayes has the quickest training time at 0.005 seconds.

In conclusion, Fidelity Fit is the best option if testing time is an important
consideration because of its greatly reduced training time
and very good accuracy. However, Linear SVM, Ada Boost, or Naive Bayes
can be taken into consideration since they offer similar accuracies, and they still have
rather quick testing times.
It is crucial to remember that the optimal classifier can change depending on
the dataset and particular issue you are attempting to solve. To ensure the
reliability of the results, it is advised to further analyze the classifiers
using cross-validation methods and study the correlation between their performances
and the quality of the qubits.

\section{Outlook}

In this proceedings we have considered a specific characterization experiment which
is performed when calibrating QPUs: the construction of a model to perform the
classification between the ground state and the excited state.
We have compared the performance of several classifiers which have been trained
on data acquired by real quantum processors at the Quantum Research Center of
Technology Innovation Institute (TII) in Abu Dhabi.
The different models
show a comparable performance when it comes to accuracy, testing time and training time.
However, due to its simplicity we believe that the Fidelity Fit seems to be the
best model for this particular dataset.

For future developments it would be interesting to perform the same study
for three levels quantum system, \emph{qutrits}. This could play a significant role
in increasing the fidelity of two-qubit gates.
Another interesting topic would be to see how the performance of the different
models changes by looking at different qubits on the same chip, which will be particularly
useful to perform noise aware classification.

\section{Acknowledgements}

The authors thank TII's Quantum Research Center for its support and Alessandro
Candido for his help implementing the benchmarking.
%
%
%
\bibliography{ref}
\end{document}